\documentclass[sigconf]{acmart}
\AtBeginDocument{%
  }

\setcopyright{none}
\copyrightyear{}
\acmYear{}
\acmDOI{}
\acmConference[Exploring the Participatory Turn in Citizen-Centred AI, an IUI'26 Workshop]{Exploring the Participatory Turn in Citizen-Centred AI, a IUI'26 Workshop}{March 23, 2026}{Paphos, Cyprus}
\acmISBN{}

\usepackage{mdframed}
\usepackage{graphicx}
\begin{abstract}
This workshop paper examines challenges in designing agentic AI systems from a citizen-centric perspective. Drawing on three participatory workshops conducted in 2025 with members of the general public and cross-sector stakeholders, we explore how societal values and expectations shape visions of future AI agents. Using constructive design research methods, participants engaged in storytelling and lo-fi prototyping to reflect on potential community impacts. We identify three key challenges: enabling meaningful and sustained public engagement, establishing a shared language between experts and lay participants, and translating speculative participant input into implementable systems. We argue that reflexive, long-term participation is essential for responsible and actionable citizen-centric AI development.
\end{abstract}

\begin{document}

\title{Three Lessons from Citizen-Centric Participatory AI Design}

\author{Eike Schneiders}
\affiliation{%
  \institution{The University of Southampton}
  \city{Southampton}
  \country{United Kingdom}}
\email{eike.schneiders@soton.ac.uk}

\author{Sarah Kiden}
\affiliation{%
  \institution{The University of Southampton}
  \city{Southampton}
  \country{United Kingdom}}
\email{sk3r24@soton.ac.uk}

\author{Beining Zhang}
\affiliation{%
  \institution{The University of Southampton}
  \city{Southampton}
  \country{United Kingdom}}
\email{b.zhang@soton.ac.uk}

\author{Bruno Rafael Queiros Arcanjo}
\affiliation{%
  \institution{The University of Southampton}
  \city{Southampton}
  \country{United Kingdom}}
\email{brqa1a24@soton.ac.uk}

\author{Zhaoxing Li}
\affiliation{%
  \institution{The University of Southampton}
  \city{Southampton}
  \country{United Kingdom}}
\email{zhaoxing.li@soton.ac.uk}

\author{Ezhilarasi Periyathambi}
\affiliation{%
  \institution{The University of Southampton}
  \city{Southampton}
  \country{United Kingdom}}
\email{ep2n24@soton.ac.uk}

\author{Vahid Yazdanpanah}
\affiliation{%
  \institution{The University of Southampton}
  \city{Southampton}
  \country{United Kingdom}}
\email{V.Yazdanpanah@soton.ac.uk}

\author{Sebastian Stein}
\affiliation{%
  \institution{The University of Southampton}
  \city{Southampton}
  \country{United Kingdom}}
\email{s.stein@soton.ac.uk}

\renewcommand{\shortauthors}{Schneiders et al.}

\begin{teaserfigure}
  \includegraphics[width=\linewidth, trim=0 1cm 0 4cm, clip]{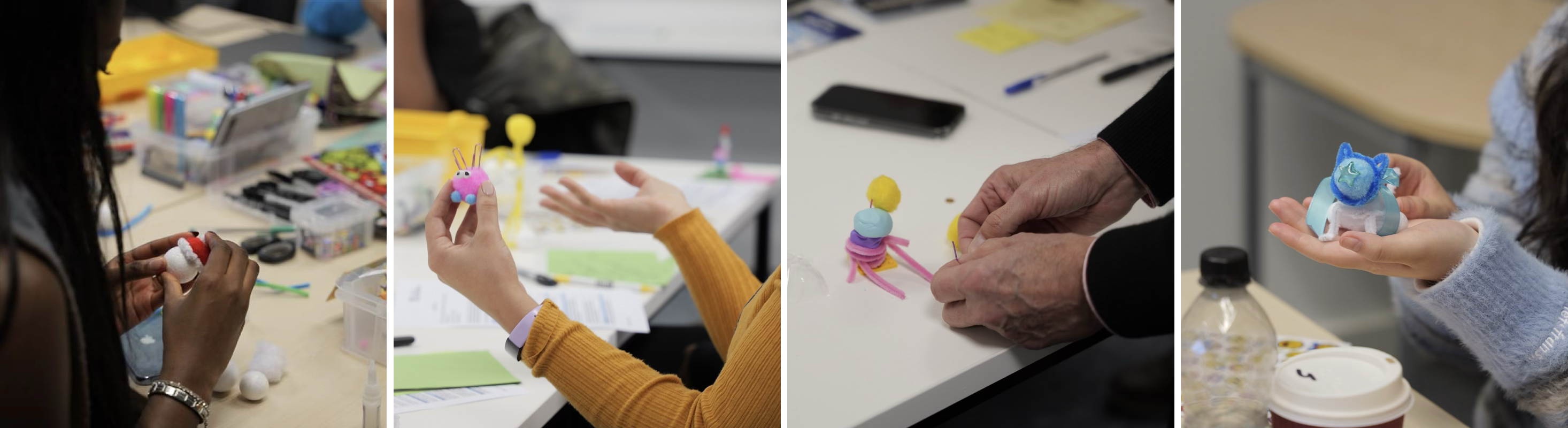}
  \caption{Impressions from participants' lo-fi prototypes of AI agents created during participatory WS1/WS2.}
  \label{fig:teaser}
\end{teaserfigure}

\maketitle

\section{Context}\label{sec:Context}
This workshop paper provides a brief introduction to key challenges when designing agentic AI-based systems using a citizen-centric perspective. The challenges identified are the result of three participatory workshops\footnote{The workshop series was conducted as part of the Citizen-Centric Artificial Intelligence System (CCAIS) project (https://www.ccais.ac.uk)\cite{stein2025multiagent}.} conducted in 2024 and 2025 (N~=~8, N~=~6, N~=~43). The workshops aimed at investigating visions, perceptions, areas of interest, and the core values which members of the general public hold in relation to AI agents. The first two workshops were aimed at members of the general public (e.g., legal experts, students, priests, journalists, retirees) while the third workshop focused on engaging a broad range of cross-sector stakeholders (e.g., government officials, AI experts, industry experts, and NGOs). In all three workshops, we emphasised the inclusion of diverse voices in the research team, as well as amongst participants, thereby eliciting a diverse set of future visions and directions for the development of AI agents~\cite{wooldridgeIntroductionMultiAgentSystems2002}. 

Inspired by constructive design research~\cite{koskinenDesignResearchPractice2011,wensveenConstructiveDesignResearch2018} the first two workshops (WS1/2) focused on participant engagement in a number of activities ranging from story completion (e.g., ``I woke up this morning and something [extraordinary / awful] had happened with the AI agent. It had ... '') to the creation of lo-fi prototypes using various materials (e.g., art/office supplies, Lego, Play-Doh, magazine cutouts, stickers, scrap paper). Finally, participants would contextualise and provide depth to their agents by narrating potential impacts, both positive and negative, these might have on the communities they were envisioned for.

The participants in workshop three (WS3) (divided into five groups) were given a design workbook, a tangible output created from the insights of WS1/2, as well as two to three artefacts (see Figure~\ref{fig:teaser}). The groups were asked to place their assigned artefacts on a timeline (between 2025 and 2035) estimating how far into the future they thought the agents could be developed and deployed. Furthermore, they were asked to assess the feasibility (i.e., if they thought the artefacts would work in real life), any suggested improvements, resources required from government and industry to make them reality, opportunities, challenges and limitations the artefacts presented. Lastly, we inquired about the requirements to current infrastructure to support the realisation of the presented visions.

\section{Challenges for Citizen-Centric AI Design}
This section will outline three challenges we encountered during the workshop series presented above. We will further highlight how we addressed these, what worked or did not work, and what lessons we learned to further improve on our citizen-centric practices.




\subsection{Meaningful Engagement is Difficult}
To prevent the risk of `participatory design,' `co-design,', or `citizen-centric design' to lose all meaning, we need to ensure meaningful engagement with individuals and communities, thereby preventing practices such as participation washing~\cite{Sloanne:2022:ParticipationWashing}. To ensure this, it is important that participation of communities, and the public at large, is long-term and not just during the evaluation. As briefly described in Section~\ref{sec:Context}, we attempted to ensure this through continuous engagement with the end-user of the envisioned agentic systems, i.e., the general public. That being said, this long term engagement does present its own challenges. Firstly, recruitment is always difficult. How do we ensure that we have diverse, across multiple dimensions, representation in the room? How do we ensure that participants see the outcomes of their participation? How do we ensure that participants meaningfully benefit from participation (beyond an Amazon voucher).

In the here described workshop series, each workshop consisted of a new set of participants. While this increases the diversity of voices, it comes at the cost of long-term engagement of the same participants. For instance, participants in WS1 might have benefited from the discussions surrounding their created artefacts in WS3. For the future, we believe it might be beneficial to ensure that participants are included in the entire process and not just in part of it. This long term involvement throughout the entire design process from ideation to potential implementation, might be greatly rewarding to participants as it might increase their appreciation for the difficulties when designing AI-based systems, thereby contributing to their AI literacy~\cite{Long:2020:AILiteracy}.

\subsection{Do we even Speak the same Language?}
As experts and lay people alike are affected by the proliferation of AI in society, we believe it is increasingly important to involve laypeople from the onset. However, this raises a crucial question we encountered in WS1/2: \textit{Do we speak the same language?} By language, we do not literally refer to the our language (e.g., English, Sámi, or Swahili), but our use and understanding of highly context and domain specific language. What is an \textit{`Agent'}? What do we mean by \textit{`Artificial Intelligence'}? Is a series of nested \texttt{If-statements} an example of AI? Is the often polarising view on AI, as presented by the mainstream media (doomerism or utopianism) really the only two options?

To ensure everyone in the room, from law student to retiree, had a common understanding of what AI is, what it might lead to, and how it might affect us, we began our two laypeople workshops (WS1/2) with a brief introduction to AI and agentic systems. We highlighted current examples, both positive and negative, and presented potential future implications of AI. However, this does not come without risk. How do we ensure that we, as researchers, do not contribute to the confusion by nudging (or even inadvertently biasing) participants due to our own perceptions? With the team comprising researchers from a Computer Science department, are we really objective? No. Of course not. So, what can we do about it? Like all qualitative data analysis (some would argue \textit{all} data analysis) a certain degree of reflexivity is crucial. We might not be unbiased and truly objective, after all, we are the ones who decided what we introduce to participants, however, it is important that we are aware of this. This awareness of our own biases, allows us to actively work on minimising the nudging of participants into a specific direction.

\subsection{Translating Participant Input into Implementable Systems}
The last challenge we wish to highlight in this short position paper, is the challenge of translation. Following WS1/2, participants had created 14 artefacts, each with their own context, use-cases, and stories attached. This now raises the question: \textit{How do we translate these into actionable insights which can inform future AI development?} To address this, we conducted WS3. Workshop 3 focused on connecting with numerous experts within the field---from government to members of industry. This allowed us to assess, or at least generate initial insights on, the feasibility of the proposed systems. What might or might not work? What would be needed to make artefact X reality? Do we (i.e., society) need to invest in human skills, infrastructure, research to make something happen? 

Going forwards, it is important to establish structured pathways for iteratively refining these laypeople-generated ideas with domain experts. This could involve pairing each artefact with targeted feasibility studies, prototyping efforts, and stakeholder feedback loops. By systematically mapping participant visions onto technical constraints (i.e., what \textit{can} be done), ethical considerations (i.e., what \textit{should} be done), and societal needs, we can ensure that the creativity and insights generated in early workshops are not lost but instead inform actionable design directions. Ultimately, this approach bridges the gap between speculative ideas and implementable systems, benefiting AI development that is both innovative and citizen-centric.

\begin{acks}
This work was supported through an EPSRC Turing AI Fellowship (EP/V022067/1), the EPSRC FEVER project (EP/W005883/1) and by Responsible Ai UK (EP/Y009800/1).
\end{acks}

\bibliographystyle{ACM-Reference-Format}
\bibliography{references}

\end{document}